\documentclass[a4paper,10pt]{article}
\usepackage{graphicx}
\usepackage[utf8]{inputenc}
\usepackage{amsmath}
\usepackage{subfig}
\usepackage{soul}
\usepackage{units}
\usepackage{xfrac}
\usepackage{float}
\usepackage{hyperref}
\usepackage{authblk}

\title{Generalized Controlled Quantum Secure Communication using GHZ-like state }
\author[]{Jemy Geordy\thanks{jemygeordy@gmail.com}}
\author[]{Dintomon Joy\thanks{dintomonjoy@cusat.ac.in}}
\author[]{Sabir M\thanks{msr@cusat.ac.in}}
\affil[]{\textit{\normalsize{Department of Physics, Cochin University of Science and Technology, \\ \textit{Kochi - 682 022, India}}}}

\begin{document}

\maketitle
\begin{abstract}
We present a novel scheme for controlled quantum secure communication (CQSC) using GHZ-like state.
  In this scheme, a trusted controller assists the users for achieving secure transmission of data between them.
  The dense coding technique is exploited  to increase the qubit efficiency and we  show that the
  proposed scheme is immune against common eavesdropping attacks.  We show how the method can be  generalized
  to include N controllers using sets of GHZ-like states shared among the controllers and users. In this
   process entanglement swapping and single particle   measurements are performed by the controllers to relax their
    controls and the users get entangled in Bell states.
\end{abstract}

\section{Introduction}

Quantum Cryptography has  emerged as area of research in recent
times. It is recognized that quantum processors can supersede
current classical cryptographic protocols and offer highly secure
cryptographic protocols for quantum communication. In $1984$,
Bennett and Brassard\cite{QKD} proposed the  pioneering work in
Quantum Key Distribution (QKD), popularly known as BB84 protocol.
This work  motivated the scientific community to exploit the
intriguing features of  of quantum mechanics to get more secure and
efficient cryptographic protocols. In QKD, initially a secret
quantum key is established between communicating parties and then
this key is used for encryption and decryption of the message which
is sent classically by Alice(sender) to Bob(receiver). With further
developments in quantum cryptography, two other kinds of protocols
were proposed. These are the quantum secure direct
communication(QSDC)\cite{pingpong, QSDC, QSDC2} scheme and the deterministic
secure quantum communication(DSQC)\cite{DSQC, teleportation, epr} scheme. Unlike QKD, the
QSDC and DSQC techniques do not require previously shared key for
communication:  the  secret messages are communicated directly.
However, there is a slight difference between QSDC and DSQC. The later requires at least one bit of classical information to
decode each qubit of the secret message and in the case of QSDC there is
no requirement of any additional classical bit to decode the secret
message\cite{Pathak, hassanpour, pathak2015}. Different QSDC and
DSQC schemes have been proposed employing different entangled states
 such as EPR pairs\cite{epr, epr2}, GHZ states\cite{DSQC, ghz}, GHZ-like states\cite{Pathak, lidong}, W states\cite{w, w2} etc.  Moreover,
there are schemes employing different methods such as  dense coding\cite{QSDC2},
entanglement swapping\cite{DSQC}, teleportation\cite{teleportation}  etc for transmitting the secret
message.

Recently, another kind of quantum communication scheme with the
inclusion of a controller  who controls the whole communication has
been put forward by many research
groups\cite{hassanpour, pathak2015, lidong, multiparty}.
In this scheme, designated as Controlled Quantum Secure Direct
Communication(CQSDC), the co-operation  of the controller is essential
for the successful communication between the legitimate users.
 In practical situations all users prefer a smaller system than the service provider. Therefore,
 it is meaningful to have a controller to prepare the state when it is needed for communication
 and have a control over the communication without the capability of accessing the secret message.
 In a recent work, Pathak\cite{pathak2015} pointed out that, since in the proposed methods
 an additional classical bit from the controller  is required to decode the secret message,
  it is more appropriate  to name this as  CDSQC (Controlled
  Secure Direct Quantum Communication) instead of CQSDC.

    In this paper we propose a new controlled quantum communication  protocol
    making use of GHZ-like states\cite{Yang, Anindita} and dense coding\cite{Bennett}. This is achieved by exploiting  a unique GHZ-like
    state: the entanglement between two particles remain even after a
    measurement is made on one of the particles\cite{hassanpour, Chao}. Since, in the proposed scheme the
    communication between the legitimate users start only after receiving a classical bit
   from the controller, we designate our protocol as Controlled Quantum Secure
    Communication(CQSC), eventhough the communication as such is purely QSDC. The generalization of our
     protocol to N number of controllers is presented, where the controllers perform  entanglement swapping also.

This paper is organized as follows. In section 2, we describe the
CQSC using a GHZ-like state and dense coding. In section 3 we extend
our scheme to CQSC with two controllers and in section 4 with three
controllers. In section 5 the method is generalized to the case of N
number of controllers (MCQSC). Analysis of the security and
efficiency of the proposed CQSC protocol is given in section 6. Our
conclusions are presented in section 7.
\section{Controlled Quantum Secure Communication using a GHZ-like state}

A GHZ-like state can be prepared from a single qubit state, Bell
pair with a controlled-NOT operation as given
 in references\cite{Pathak, hassanpour}. The set of eight orthonormal GHZ-like states\cite{hassanpour} is given by,
\begin{eqnarray}
\left|\zeta_{000}\right\rangle = \frac{1}{\sqrt{2}}[\left|\phi^{+}\right\rangle\left|0\right\rangle + \left|\psi^{+}\right\rangle\left|1\right\rangle],\nonumber \quad \left|\zeta_{001}\right\rangle = \frac{1}{\sqrt{2}}[\left|\phi^{+}\right\rangle\left|1\right\rangle + \left|\psi^{+}\right\rangle\left|0\right\rangle]\\ \nonumber
\left|\zeta_{010}\right\rangle = \frac{1}{\sqrt{2}}[\left|\phi^{+}\right\rangle\left|0\right\rangle - \left|\psi^{+}\right\rangle\left|1\right\rangle],\nonumber    \quad\left|\zeta_{011}\right\rangle = \frac{1}{\sqrt{2}}[\left|\phi^{+}\right\rangle\left|1\right\rangle - \left|\psi^{+}\right\rangle\left|0\right\rangle]\\\nonumber
\left|\zeta_{100}\right\rangle = \frac{1}{\sqrt{2}}[\left|\phi^{-}\right\rangle\left|0\right\rangle + \left|\psi^{-}\right\rangle\left|1\right\rangle],\nonumber \quad\left|\zeta_{101}\right\rangle = \frac{1}{\sqrt{2}}[\left|\phi^{-}\right\rangle\left|1\right\rangle + \left|\psi^{-}\right\rangle\left|0\right\rangle]\\\nonumber
\left|\zeta_{110}\right\rangle = \frac{1}{\sqrt{2}}[\left|\phi^{-}\right\rangle\left|0\right\rangle - \left|\psi^{-}\right\rangle\left|1\right\rangle],\nonumber \quad\left|\zeta_{111}\right\rangle = \frac{1}{\sqrt{2}}[\left|\phi^{-}\right\rangle\left|1\right\rangle - \left|\psi^{-}\right\rangle\left|0\right\rangle]
\end{eqnarray}

where, $ \left|\phi^{\pm}\right\rangle $ and
$\left|\psi^{\pm}\right\rangle$ represents the Bell states.
\begin{eqnarray*}
\left|\phi^{\pm}\right\rangle = \frac{1}{\sqrt{2}}[\left|00\right\rangle \pm \left|11\right\rangle], \quad\left|\psi^{\pm}\right\rangle = \frac{1}{\sqrt{2}}[\left|01\right\rangle \pm \left|10\right\rangle]\\
\end{eqnarray*}

In this protocol, we consider Alice as the sender who wants to send secret message to Bob the
 receiver and Charlie controls the communication. The proposed CQSC consists of the following steps,

\begin{figure}[ht]
\centering
\subfloat[][]{{\includegraphics[width=2.5cm]{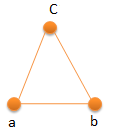}}}%
\qquad
\subfloat[][]{{\includegraphics[width=2.5cm]{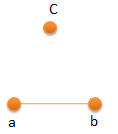}}}%
\caption[]{CQSC with one controller, (a) initially shared GHZ-like state among the controller and the communicators Alice and Bob, (b) after the Z basis measurement by the controller C, the particles \textit{a} and \textit{b} become entangled.}%
\label{1QSC}%
\end{figure}

\textbf{CQSC 1}: Charlie prepares N number of GHZ-like state of the form,
\begin{equation}
\label{eqn 1}
\left|\zeta_{000}\right\rangle = \frac{1}{2}[\left|000\right\rangle + \left|011\right\rangle + \left|110\right\rangle + \left|101\right\rangle]_{abc}
\end{equation}

This can also be written as,
\begin{equation}
 \label{bell2} \left|\zeta_{000}\right\rangle =
[\frac{(\left|\phi^{+}\right\rangle\left|0\right\rangle +
\left|\psi^{+}\right\rangle\left|1\right\rangle)}{\sqrt{2}}]_{abc}
\end{equation}
in the Z basis and as
\begin{equation}
\label{bell}
\left|\zeta_{000}\right\rangle = \frac{1}{2}[\left|+++\right\rangle + \left|---\right\rangle]_{abc}\\
\end{equation}
in the X basis.

 Charlie, now, has in his possession a sequence of N number of GHZ-like state
$[P_{1}(a)P_{1}(b)P_{1}(c),P_{2}(a)P_{2}(b)P_{2}(c),..,P_{N}(a)P_{N}(b)P_{N}(c)]$,
where the a,b,c represents three particles in the GHZ-like state and
the subscript represents the order. Charlie divides the sequence
into three, `a' sequence : $[P_{1}(a)P_{2}(a)P_{3}(a),..,P_{N}\\(a)]$,
`b' sequence: $[P_{1}(b)P_{2}(b)P_{3}(b),..,P_{N}(b)]$ and `c'
sequence: $[P_{1}(c)P_{2}(c)P_{3}(c)\\,..,P_{N}(c)]$. Charlie then
sends the `a' sequence to Alice,  `b' sequence to Bob and retains
the `c' sequence with himself. The communicating parties
are already aware of the shared GHZ-like state.

\textbf{CQSC 2}: The controller Charlie randomly selects a
sufficiently large subset of particles from his `c' sequence for the
first eavesdrop checking. These particles are
 then  measured in either Z or X basis at random. Later the position of the sample
  particles and the basis chosen for measurement are announced via a public channel.
 Alice and Bob choose the same basis to measure the corresponding particle in their sequence.
 When Alice and Bob tell their measurement results to Charlie, he can find out the error rate.
  From equation~\ref{eqn 1}, if Charlie gets $\left|0\right\rangle_c$ in the Z basis measurement,
  then the results of Alice and Bob are correlated and if Charlie gets $\left|1\right\rangle_c$,
 then their results are anti-correlated. Similarly results corresponding to Charlie's measurement
  in X basis are related as given in equation \ref{bell}. They continue to the next step
 if the error rate is low, otherwise they abort the communication and start from the first step on a later occasion.

\textbf{CQSC 3}: After the first eavesdrop check, the controller
performs Z-basis measurement on the
 remaining particles in the `c' sequence. This measurement allows the controller to disentangle all
 particles in the `c' sequence from the sequences `a' and `b'. At this stage Alice and Bob share Bell pairs according
 to equation \ref{bell2} as  is pictorially shown in figure \ref{1QSC}. However, Charlie shares
  his measurement result only with Bob through a classical channel. This allows the receiver to have
   an advantage over eavesdropper. Now Bob and Charlie alone know which Bell pair is shared between Alice
 and Bob, since they already have information about the initially shared GHZ-like state.

\textbf{CQSC 4}: To enhance the security, Bob does any one of the four unitary operations
 I, $\sigma_{x}, \sigma_{z}, i\sigma_{y}$ randomly on his `b' sequence  $[P_{1}(b)P_{2}(b)P_{3}(b),..,P_{N}(b)]$
 remaining after the first eavesdrop check. These operations transform the Bell states shared
  between Alice and Bob accordingly. This transformation protects the information about
  the shared entanglement between the users also from controller's knowledge. Now, only Bob
  knows which are the shared Bell pairs. Bob informs Alice that he is ready to receive the message.
\begin{table}[!ht]
\centering
\begin{tabular}{l l}
\hline\noalign{\smallskip}
Unitary operation  &  Encoded message\\
performed by Alice  &  \\
\noalign{\smallskip}\hline\noalign{\smallskip}
$I$   &  00\\
$\sigma_{x}$   &  01\\
$\sigma_{z}$   &  10\\
$i\sigma_{y}$  &  11\\
\noalign{\smallskip}\hline\noalign{\smallskip}
\end{tabular}
\caption{Unitary operations and the corresponding encoded message}\label{tab:1}
\end{table}

\textbf{CQSC 5}: After receiving the confirmation from Bob, Alice encodes her secret
 message to the `a' sequence $[P_{1}(a)P_{2}(a)P_{3}(a),..,P_{N}(a)]$ remaining with her after
 the first eavesdrop check. For encoding the secret message Alice uses one of the
 four unitary operations \textit{I}, $\sigma_{x}, \sigma_{z}, i\sigma_{y}$ and each
 one corresponding to two bit of information as shown in table~\ref{tab:1}. Before
 encoding the message, Alice inserts sufficient decoy photons in her sequence in order
 to confuse the eavesdropper in decoding the secret message. This process
would  improve the security of the protocol. Now Alice sends her
encoded \textbf{`a'} sequence
   along with the decoy photons to Bob through a quantum channel as in the case of dense
   coding\cite{Bennett}.

\textbf{CQSC 6}: After receiving the encoded message sequence from Alice, Bob sends a confirmation
signal back to Alice. On receiving the confirmation message from Bob, Alice announces the positions
 and the basis of the decoy photons. Bob  will measure the decoy photons in the respective basis and
 tells his result to Alice so that she can calculate the error rate. Now, Bob has both the
 qubits of the EPR pair. On these transformed EPR pairs, Bob performs Bell basis measurements.
  With the knowledge of the unitary operation that he has earlier done on the `b' sequence
 before the encoding process and with the Bell measurement results, Bob can obtain
  the secret message sent by Alice. Since Bob initially changed the state, Eve(the eavesdropper)
  will get nothing even if she eavesdrops.

For an example, consider that Alice wants to send a 6 bit message, say, 100101 to Bob.
Let $\left|\zeta_{000}\right\rangle = [\frac{(\left|\phi^{+}\right\rangle\left|0\right\rangle + \left|\psi^{+}\right\rangle\left|1\right\rangle)}{\sqrt{2}}]_{ABC}$
be the GHZ-like state shared among the two users Alice and Bob, and the controller Charlie. They
need three sets of GHZ-like state after the first eavesdrop check to send this 6 bit of information.
 Alice, Bob and Charlie share the set of three GHZ-like state as elaborated in CQSC:1 of our protocol.
 After the first eavesdrop check as per CQSC:2, Charlie measures his C-sequence as explained in CQSC: 3
  and informs Bob about his measurement. Let Charlie's measurement results be $\left|0\right\rangle, \left|1\right\rangle$
  and $ \left|0\right\rangle$ so that the shared Bell states between Alice and Bob are
  $ \left|\phi^{+}\right\rangle, \left|\psi^{+}\right\rangle$ and $\left|\phi^{+}\right\rangle$ respectively.
  Then Bob randomly select any three unitary operations as explained in CQSC:4, which makes him the
  only user who knows the shared Bell state. For instance, say, Bob selects $\sigma_{x}$, $I$ and $\sigma_{z}$
  respectively to operate his B sequence, then the shared Bell states transform to
  $ \left|\psi^{+}\right\rangle, \left|\psi^{+}\right\rangle$ and  $\left|\phi^{-}\right\rangle $.
   Now Alice encodes her message on her A sequence using unitary operations as explained in CQSC: 5.
    To send 100101, she selects $\sigma_{z}$, $\sigma_{x}$ and $\sigma_{x}$ respectively to operate on
    her qubits. Alice's operation transform the Bell pairs to $ \left|\psi^{-}\right\rangle, \left|\phi^{+}\right\rangle$
    and $\left|\psi^{-}\right\rangle $. Alice sends her encoded sequence to Bob after inserting decoy photons
    to check error rate as explained in CQSC: 6.  After removing the decoy photos Bob does Bell measurement on
    the Bell pair. His measurement result for this particular example is 110011. With the knowledge of shared
     Bell pair before encoding the message, Bob can analyse the operation done by Alice and hence
      gets the secret message 100101.

\section{CQSC with two controllers}

 Here we elaborate our protocol on how to realise the communication between Alice and Bob with the help of two trusted controllers $C_{1}$ and $C_{2}$. It requires two sets of GHZ-like states to achieve this two-controller communication. The CQSC scheme is explained in following steps,
\begin{figure}[ht]
\centering
\subfloat[][]{{\includegraphics[width=3.5cm]{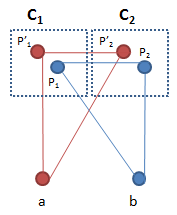}}}%
\qquad
\subfloat[][]{{\includegraphics[width=3.6cm]{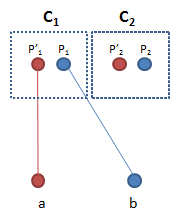}}}%
\qquad
\subfloat[][]{{\includegraphics[width=4cm]{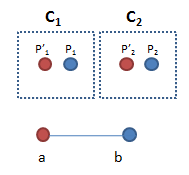}}}%
\caption[]{CQSC with two controllers, (a) initially shared GHZ-like states among the two controllers and the communicators, (b) the controller $C_{2}$ measures his two particles $p_{2}$ and $p'_{2}$ in Z basis, this changes the particles $p'_{1}$ and $a$ , $p_{1}$ and $b$ to Bell pairs, (c) the controller $C_{1}$ swaps his particles $p_{1}$ and $p'_{1}$, which turns the particles \textit{a} and \textit{b} to Bell pair.}%
\label{2QSC}%
\end{figure}

\textbf{2CQSC 1:} The controller $C_{1}$ prepares N number of two sets of GHZ-like state, in the states $ \left|\zeta_{1}\right\rangle$ and $\left|\zeta_{2}\right\rangle$.
\begin{eqnarray}
\label{2cbell}
\left|\zeta_{1}\right\rangle = [\frac{(\left|\phi^{+}\right\rangle\left|0\right\rangle + \left|\psi^{+}\right\rangle\left|1\right\rangle)}{\sqrt{2}}]_{p'_{1}ap'_{2}}\\
\label{2cbell2}
\left|\zeta_{2}\right\rangle = [\frac{(\left|\phi^{+}\right\rangle\left|0\right\rangle + \left|\psi^{+}\right\rangle\left|1\right\rangle)}{\sqrt{2}}]_{p_{1}bp_{2}}
\end{eqnarray}
The first set of GHZ-like state is composed of three particles $(p'_{1}$, \textit{a} and $p'_{2})$ and those in second set of GHZ-like state are $(p_{1}$, \textit{b} and $p_{2})$. The ordered sequence of the two set of GHZ-like state is given by $[P_{1}(p'_{1})P_{1}(a)P_{1}(p'_{2}), P_{2}(p'_{1})P_{2}(a)P_{2}(p'_{2})\\,...,P_{N}(p'_{1})P_{N}(a)P_{N}(p'_{2})]$ and $[P_{1}(p_{1})P_{1}(b)P_{1}(p_{2}), P_{2}(p_{1})P_{2}(b)P_{2}(p_{2}),..,P_{N}(p_{1})\\P_{N}(b)P_{N}(p_{2})]$. The controller $C_{1}$ divides the two set of GHZ-like states into six sequences: $[p'_{1}, a, p'_{2}, p_{1}, b, p_{2}]$ 
and distribute the sequences among the users and controllers as follows. Sequences $p_{2}$: $[P_{1}(p_{2})P_{2}(p_{2}),..,P_{N}(p_{2})]$, $p'_{2}$: $[P_{1}(p'_{2})P_{2}(p'_{2})\\,.,P_{N}(p'_{2})]$ are given to the controller $C_2$ and sequences \textit{a}: $[P_{1}(a)P_{2}(a),..,P_{N}(a)]$, \textit{b}: $[P_{1}(b)P_{2}(b),..,P_{N}(b)]$ are given to the users Alice and Bob respectively. The remaining two sequences $c_{1}$: $[P_{1}(p_{1})P_{2}(p_{1}),..,P_{N}(p_{1})]$ and $p'_{1}$: $[P_{1}(p'_{1})P_{2}(p'_{1}),..,\\P_{N}(p'_{1})]$ are retained by $C_1$. The controllers and the two communicating parties are aware of the shared GHZ-like state.

\textbf{2CQSC 2:} The controller $C_2$ initiates the first eavesdrop check by choosing a large subset of particles from the sequence $p_{2}$ and $p'_{2}$ and measures the particles randomly in X or Z basis. Then $C_2$ announces the measurement basis and the result in a public channel. Based on the information published by $C_2$ Alice and Bob performs measurement on the corresponding particles in their sequence in same basis as chosen by $C_2$. Alice and Bob share their results to the controller $C_{1}$ and in turn $C_1$ measures the corresponding particles in $p_{1}$ and $p'_{1}$ sequence in the same basis as choosen by $C_2$. By comparing the measurement results, $C_1$ can find the error rate. If the error rate is below the threshold, they continue their communication, otherwise $C_1$ informs others to abort the communication and start the whole process from the beginning.

\textbf{2CQSC 3:} After the eavesdrop check, the controller $C_{1}$ insists $C_{2}$ to measure the two particles sequences $p_{2}$ and $p'_{2}$ in Z basis. As a result of these measurements entanglement is established between pairs ($ p'_1, a $) and ($ p_1, b$) according to equation \ref{2cbell} and \ref{2cbell2}, as shown in figure:~\ref{2QSC}(b). After the public announcement of measurement results by controller $C_2$, $C_1$ performs bell measurement on two sequence  $p_{1}$ and  $p'_{1}$ in order. This measurement allows entanglement to be established between sequences \textit{a} and \textit{b}. The Bell measurement result of  $C_{1}$ is shared with the receiver Bob and then the communication proceeds as in one controller case as explained in previous section, in step \textbf{CQSC 4}. The possible measurement outputs of the two controllers and the shared Bell pair between the sender and receiver is given in table:~\ref{tab:2}. 

\begin{table}[H]
\centering
\begin{tabular}{l l l}
\hline\noalign{\smallskip}
$C_{2}$   &  $C_{1}$  &  Shared bell pair  \\
($p'_{2}$ $p_{2}$) & ($p'_{1}$ $p_{1}$)  &  between Alice and Bob \\
\noalign{\smallskip}\hline\noalign{\smallskip}
0 0  & 0 0  & $\left|\phi^{+}\right\rangle$\\
  & 0 1 &  $\left|\psi^{+}\right\rangle $\\
  &  1 0  &  $\left|\phi^{-}\right\rangle$\\
    &  1 1 &  $\left|\psi^{-}\right\rangle$\\
\noalign{\smallskip}\hline\noalign{\smallskip}
0 1    &  0 0 &  $\left|\psi^{+}\right\rangle$\\
        &  0 1 &  $\left|\phi^{+}\right\rangle$\\
        &  1 0 &  $\left|\psi^{-}\right\rangle$\\
            &  1 1 &  $\left|\phi^{-}\right\rangle$\\
\noalign{\smallskip}\hline\noalign{\smallskip}
1 0 & 0 0 &  $\left|\psi^{+}\right\rangle$\\
        &  0 1 &  $\left|\phi^{+}\right\rangle$\\
        &  1 0 &  $\left|\psi^{-}\right\rangle$\\
            &  1 1 &  $\left|\phi^{-}\right\rangle$\\       
\noalign{\smallskip}\hline\noalign{\smallskip}
1 1 &  0 0 &  $\left|\phi^{+}\right\rangle$\\
 &  0 1 &  $\left|\psi^{+}\right\rangle$\\
 &  1 0 &  $\left|\phi^{-}\right\rangle$\\
 &  1 1 &  $\left|\psi^{-}\right\rangle$\\   
\noalign{\smallskip}\hline\noalign{\smallskip}
\end{tabular}\caption{CQSC with two controllers}\label{tab:2}
\end{table} 

\section{CQSC with three controllers}

In the case of three controller QSC, we consider $C_1$, $C_2$ and $C_3$ as the trusted controllers. Here again we need only two sets of GHZ-like state to realise three-party Controlled quantum secure communication between legitimate users. The 3CQSC scheme is explained as follows.

\begin{figure}[ht]
\centering
\subfloat[][]{{\includegraphics[width=4.5cm]{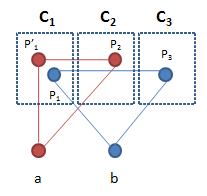}}}%
\qquad
\subfloat[][]{{\includegraphics[width=4.6cm]{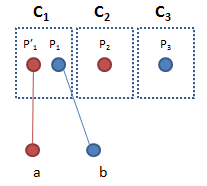}}}%
\qquad
\subfloat[][]{{\includegraphics[width=4.5cm]{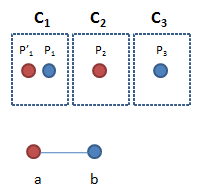}}}%
\caption[]{CQSC with three controllers, (a) initially shared GHZ-like states among the three controllers and the communicators, (b) the controller $C_{2}$ and $C_{3}$ measures their particles $p_{2}$ and $p_{3}$ in Z basis, this changes the particles $p'_{1}$ and $a$ , $p_{1}$ and $b$ to Bell pairs, (c) the controller $C_{1}$ swaps his particles $p_{1}$ and $p'_{1}$, which turns the particles \textit{a} and \textit{b} to Bell pair.}%
\label{3QSC}%
\end{figure}

\textbf{3CQSC 1:} The controller $C_{1}$ prepares N number of two sets of GHZ-like state $\{\left|\zeta_{1}\right\rangle$, $\left|\zeta_{2}\right\rangle\}$ and distributes the particles among the other controller $C_{2}$, $C_{3}$ and the users a and b. Let the two sets of GHZ-like state be,

\begin{eqnarray}
\label{3cbell}
\left|\zeta_{1}\right\rangle = [\frac{(\left|\phi^{+}\right\rangle\left|0\right\rangle + \left|\psi^{+}\right\rangle\left|1\right\rangle)}{\sqrt{2}}]_{p'_{1}ap_{2}}\\
\label{3cbell2}
\left|\zeta_{2}\right\rangle = [\frac{(\left|\phi^{+}\right\rangle\left|0\right\rangle + \left|\psi^{+}\right\rangle\left|1\right\rangle)}{\sqrt{2}}]_{p_{1}bp_{3}}
\end{eqnarray}
where, $(p'_{1}$, \textit{a} and $p_{2})$ represents particles that belong to the first set of GHZ-like state $\left|\zeta_{1}\right\rangle$ and particles $(p_{1}$, \textit{b} and $p_{3})$ belong to the second set $\left|\zeta_{2}\right\rangle$. Now the controller $C_1$ divides the N number of GHZ-like states in to six sequences: $\{p'_{1}$, \textit{a}, $p_{2}$, $p_{1}$, \textit{b}, $p_{3}\}$ and distributes it among the controllers and users as follows. The sequences $p_2$ : $[P_{1}(p_{2})P_{2}(p_{2}),..,P_{N}(p_{2})]$, $p_3$: $[P_{1}(p_{3})P_{2}(p_{3}),..,P_{N}(p_{3})]$ and \textit{a}:$[P_{1}(a)P_{2}(a),..,P_{N}(a)]$, \textit{b}: $[P_{1}(b)P_{2}(b),..,P_{N}(b)]$  are given to Controllers $C_2$, $C_3$ and users a, b respectively. The remaining sequences 
$c'_{1}$: $[P_{1}(p'_{1})P_{2}(p'_{1}),..,P_{N}\\(p'_{1})]$ and $p_{1}$: $[P_{1}(p_{1})P_{2}(p_{1}),..,P_{N}(p_{1})]$ are retained by the Controller $C_1$. Here the shared GHZ-like states are known to the controllers and the two communicating parties.

\textbf{3CQSC 2:} For the first eavesdrop checking, the controllers $C_{2}$ and $C_{3}$ chooses a large subset of particles with the same order from the sequence $p_{2}$, $p_{3}$ and measures the particles randomly in X or Z basis. They announce their measurement basis and results in a public channel. Using the obtained information, Alice and Bob chooses the corresponding particles with same order from the sequences $a$, $b$ and selects measurement basis same as that of $C_{2}$ and $C_3$ respectively. Their measurement results are communicated only with the controller $C_{1}$. Later, $C_1$ measures the corresponding particles in the $p'_1$, $p_1$ sequences with the same order in the basis same as that of $C_2$ and $C_3$ respectively. By comparing the results of other users, $C_1$ can calculate the error rate. If the error rate is low they proceed to the next step, otherwise they abort the communication and start from the beginning.
 
\textbf{3CQSC 3:} The controllers $C_{2}$ and $C_{3}$ are instructed to measure their particle sequences $p_{2}$ and $p_{3}$ in Z basis. The measurement of $p_{2}$ allows the Controller $C_{1}$ and the sender Alice to share Bell pairs as in equation \ref{3cbell}. Similarly the measurement of $p_{3}$ allows the Controller $C_{1}$ and the receiver Bob to share Bell pairs according to equation \ref{3cbell2}. The entanglement relation between the particles after the measurements are pictorially shown in figure:~\ref{3QSC}(b). The controller $C_{2}$ and $C_{3}$ announce their results publicly. After this announcement the controller $C_{1}$ performs bell measurement on the two sequence $p_{1}$ and $p'_{1}$ with the same order. This in turn swaps the entanglement between $C_1$ and users to only among user sequences \textit{a} and \textit{b} as shown in figure:~\ref{3QSC}(c). The Bell measurement results of $C_{1}$ is shared with the receiver Bob and now the communication proceeds as in one controller case \textbf{CQSC 4}. The possible measurement results of the three controllers and the shared Bell pair between the sender and receiver is given in table~\ref{table3Controllers} and is similar as the case of two controllers.

\begin{table}[H]
\centering
\begin{tabular}{l l l l}
\hline\noalign{\smallskip}
$C_{3}$ & $C_{2}$   &  $C_{1}$  &  Shared bell pair  \\
($p_{3}$)& ($p_{2}$) & ($p'_{1}$ $p_{1}$)  &  between Alice and Bob \\
\noalign{\smallskip}\hline\noalign{\smallskip}
0 & 0  & 0 0  & $\left|\phi^{+}\right\rangle$\\
 & & 0 1 &  $\left|\psi^{+}\right\rangle $\\
 & &  1 0  &  $\left|\phi^{-}\right\rangle$\\
  &  &  1 1 &  $\left|\psi^{-}\right\rangle$\\
\noalign{\smallskip}\hline\noalign{\smallskip}
0& 1    &  0 0 &  $\left|\psi^{+}\right\rangle$\\
    &    &  0 1 &  $\left|\phi^{+}\right\rangle$\\
     &   &  1 0 &  $\left|\psi^{-}\right\rangle$\\
       &     &  1 1 &  $\left|\phi^{-}\right\rangle$\\
\noalign{\smallskip}\hline\noalign{\smallskip}
1& 0 & 0 0 &  $\left|\psi^{+}\right\rangle$\\
    &    &  0 1 &  $\left|\phi^{+}\right\rangle$\\
      &  &  1 0 &  $\left|\psi^{-}\right\rangle$\\
       &     &  1 1 &  $\left|\phi^{-}\right\rangle$\\       
\noalign{\smallskip}\hline\noalign{\smallskip}
1 &1 &  0 0 &  $\left|\phi^{+}\right\rangle$\\
 & &  0 1 &  $\left|\psi^{+}\right\rangle$\\
 & &  1 0 &  $\left|\phi^{-}\right\rangle$\\
 & &  1 1 &  $\left|\psi^{-}\right\rangle$\\   
\noalign{\smallskip}\hline\noalign{\smallskip}
\end{tabular}\caption{CQSC with three controllers}\label{table3Controllers}
\end{table}

\section{Generalized controlled Quantum Secure Communication}

In this section, we generalize our protocol for CQSC with N trusted controllers. Let $ C_{1}, C_{2}, C_{3}... C_{N}$ denote the N controllers. The particle distribution of each GHZ-like state among the controllers and total number of GHZ-like states required for establishing the protocol can be found from the equations given below. Here, each line represents an entangled set of GHZ-like state.

\textbf{For odd number of controllers:}

\begin{equation}
\left\{p'_{\frac{N-1}{2}} ,	p'_{\frac{N-3}{2}},	p_{\frac{N+1}{2}} \right\}\\
\label{first}
\end{equation}
Equation:~\ref{first}, represents the three particles from the first set of entangled GHZ-like state. The number in subscript represents the identity of each controller and it depends on the total number of odd  controllers, N. These particles are send to different controllers based on the value in the subscript. The remaining set of GHZ-like states is distributed among the controllers according to the following pattern:
\begin{equation}
\left\{p_{\frac{N-(2m+1)}{2}}, p'_{\frac{N-(2m+5)}{2}}, p_{\frac{N+(2m+3)}{2}}\right\}\\
\end{equation}

For different values of m, the above equation generates entangled states which has to be distributed among different controllers. It terminates when the value of the middle term, $(\nicefrac{N-(2m+5)}{2})$ becomes $-1$. Therefore, $m$ varies from $m=\{0,1,2,....,\frac{N-3}{2}\}$, generating $m+1$ set of GHZ-like states or $\frac{N-1}{2}$ set. Now, by including the first set given in equation:~\ref{first}, the total number of GHZ-like states required for the odd number of controllers case becomes $\frac{N+1}{2}$. On termination, the terms $[p'_{0}, p'_{-1}]$ are replaced by [\textit{a},\textit{b}] and are send to the legitimate users Alice and Bob respectively. 

\textbf{For even number of controllers:}

\begin{equation}
\left\{p'_{\frac{N}{2}}	,	p'_{\frac{N-2}{2}},	p'_{\frac{N+2}{2}}\right\}\\
\label{second}
\end{equation}
Similar to the odd number case, equation:~\ref{second} represents the first set of GHZ-like state. The equation given below generates the remaining set of entangled states depending on the value of $m$. 
\begin{equation}
\left\{p_{\frac{N-2m}{2}}, p'_{\frac{N-(2m+4)}{2}}, p_{\frac{N+(2m+2)}{2}}\right\}\\
\end{equation}

The above sequence of equations also terminate, when the value of middle term $(\nicefrac{N-(2m+4)}{2})$ becomes $-1$. This gives the value of $m$ to  be varying from $m=\{0,1,2,....,\frac{N-2}{2}\}$ generating $m+1$ set of GHZ-like state or $\frac{N}{2}$ set. Again by including the entangled state given in equation:~\ref{second}, the total number of required entangled states can be found to be $\frac{N+2}{2}$. On termination, as in the previous case, the terms $[p'_{0}, p'_{-1}]$ are replaced by [\textit{a},\textit{b}] and are send to the legitimate users Alice and Bob respectively.                                                                                         

The controller $C_{1}$ prepares the required number of GHZ-like states and distribute it among other controllers and users. $C_{1}$ retains particle with  subscripts corresponding to his identity. After the distribution, $C_{1}$ informs the controllers possessing third particle of each GHZ-like state to measure their particle in Z basis, which makes the other two particles of corresponding GHZ-like state collapse to Bell state. $C_{1}$ informs the controllers with the first particle of each GHZ-like state to perform Bell measurement with the other particle in their possession. Then the controllers numbering from $C_{\frac{N-1}{2}}$ to $C_{1}$ for odd number of controllers and $C_{\frac{N}{2}}$ to $C_{1}$ for even number of controllers will perform bell measurement on their respective particles in the decreasing order of controller number given in subscript. After the final Bell measurement from $C_{1}$, an entanglement channel gets established between the users to enable quantum secure communication. 

\section{Security analysis and efficiency of the proposed CQSC}

In this section we are analysing the security of our scheme towards different kinds of eavesdropping attacks. The most common attacks includes; intercept-resend attack, measure-resend attack, entanglement attack etc.

\textbf{The intercept-resend attack}: In this attack, Eve creates N sets of GHZ-like state, say $\left|\zeta\right\rangle_{xyz}$. She may capture the \textit{a} and \textit{b} sequence that have been sent to Alice and Bob by Charlie in the step \textbf{CQSC 1} and replace those \textit{a} and \textit{b} sequences with her \textit{x} and \textit{y} sequences and resend it to Alice and Bob. However, on the first eavesdropping check, the presence of an eavesdropper can be found from the error rate calculated by Charlie, because of the non-correlation between C sequence of Charlie and \textit{x}, \textit{y} sequences of Eve.

Likewise, Eve may again try to intercept the encoded \textit{a} sequence during its transmission from Alice to Bob. However, in the second eavesdropping check (\textbf{CQSC 6}) the presence of Eve gets revealed by the disturbance she creates on the decoy photons. The ignorance of Eve, about the position and the basis in which the decoy photons are created, reveals the presence of Eve. 

\textbf{The measure-resend attack}: In this case, Eve may measure each particle of \textit{a} and \textit{b} sequence during its transmission from Charlie to Alice or Bob. After that she resend those particles to the corresponding users. However, this action performed by Eve will destroy the entangled GHZ-like state and the presence of Eve will get revealed in the first eavesdrop checking. Even if she captures the message encoded \textit{a} sequence, her lack of knowledge about the position and basis with which the decoy photons are created, reveals her presence in the second eavesdrop check.

\textbf{The entangle-measure attack}: In this type of attack, eavesdropper may try to entangle her ancilla qubits with the travelling qubit sequences \textit{a} and \textit{b} sent by Charlie to Alice and Bob respectively. For instance, Eve can entangle her ancilla qubits $\left|00\right\rangle_{xy}^{(n)}$ with the particles in the sequences \textit{a} and \textit{b} by two CNOT operations. Where, $n$ in the superscript represents the order of ancilla bits. The particles (\textit{a}, \textit{b}) act as control bits and (\textit{x}, \textit{y}) act as target bits. The appropriate unitary operation U is explicitly given below.

\begin{eqnarray*}
U\left|\Psi\right\rangle_{abc12} &=& CNOT_{(ax)}CNOT_{(by)}\left\{\frac{\left|\phi^{+}\right\rangle_{ab}\left|0\right\rangle_c+ \left|\psi^{+}\right\rangle_{ab}\left|1\right\rangle_c }{\sqrt{2}}\right\} \otimes\left|00\right\rangle_{xy}\\\nonumber
&=&\frac{1}{2}\{\left|00000\right\rangle + \left|01101\right\rangle + \left|11011\right\rangle + \left|10110\right\rangle\}_{abcxy}\\\nonumber
&=&  \frac{1}{\sqrt{2}}\{ (\left|\phi^{+}\right\rangle_{ab}\left|\phi^{+}\right\rangle_{xy}+\left|\phi^{-}\right\rangle_{ab}\left|\phi^{-}\right\rangle_{xy})\left|0\right\rangle_{c} \\&+& (\left|\psi^{+}\right\rangle_{ab}\left|\psi^{+}\right\rangle_{xy}+\left|\psi^{-}\right\rangle_{ab}\left|\psi^{-}\right\rangle_{xy})\left|1\right\rangle_{c} \}\\\nonumber
\end{eqnarray*}
The entanglement of Eve's particles (\textit{x}, \textit{y}) with the qubits (\textit{a}, \textit{b}) creates two additional undesired bell state results, which will increase the error in the expected correlations by 50\%. Hence in the first eavesdrop checking, Charlie can detect the presence of an eavesdropper 50\% of times. Even if Eve tries to entangle her ancilla bits with the message encoded \textit{a} sequence, Eve will get detected because of the disturbance she creates on decoy photons. Also Eve gets restricted from obtaining any useful information, because of the random unitary operations performed by Bob in his \textit{b} sequence before Alice encodes her secret message on the \textit{a} sequence. In our protocol we have supposed that all controllers are trustworthy. The proposed scheme can be made more secure by the inclusion of an identity authentication step between the users, before decoding the message.

\textbf{Efficiency:} Efficiency of the proposed protocol can be calculated using the equation\cite{liu},

\begin{equation}
\eta = \frac{c}{q + b}
\end{equation}

where c is the number of classical bits transmitted, 
q is the number of qubits used,
b is the number of classical bits exchanged.
For our scheme $c = 2, q = 3$ and $b = 1$. That is, we will get an efficiency of 50\%.

\section{Conclusion}

We have presented a controlled quantum secure communication protocol using GHZ-like state as quantum channel and employed the dense coding technique to increase the qubit efficiency. The sender Alice and receiver Bob are allowed to communicate only when the controller Charlie allows them. Bob performs random unitary operations on particle in his possession to secure the information about shared entanglement, from the controller's knowledge. Alice encodes her secret messages on particles in her possession by performing one of four different unitary operations, each corresponding to a different two bit message, and transmits it to Bob. By performing Bell measurement on particles in his possession, Bob read out the secret messages. 

Further, we have generalized our scheme to N-controllers using a set of similar GHZ-like states. In this multiparty controlled quantum secure communication, one of the controller prepares and efficiently distributes the entangled states among the other controllers. The users then require the co-operation from N controllers to begin their secret communication, which may be useful in many practical applications. We have also shown that our protocol is secure against common eavesdropping attacks and the qubit efficiency found to be 50\%, which is relatively higher than other existing QSDC schemes. Our protocol is feasible with the current technology as it requires only single particle measurements and entanglement swapping to establish secure communication channel between users. However, our protocol demands all the controllers to be trustworthy for successful communication. 

\section*{Acknowledgements}
One of the author Jemy thank Sarath R for his valuable comments on the manuscript.

\end{document}